\documentclass[12pt]{article}
\usepackage{color}           
\usepackage[a4paper,left=30mm,right=20mm,top=20mm,bottom=20mm]{geometry}
\usepackage{graphics}
\usepackage{amsmath}
\usepackage{epsfig}
\usepackage{cite}
 \newcommand{\be}{\begin{eqnarray}}
 \newcommand{\ee}{\end{eqnarray}}
 \newcommand{\beq}{\begin{equation}}
 \newcommand{\eeq}{\end{equation}}
 \newcommand{\ba}{\begin{array}{1}}
 \newcommand{\ea}{\end{array}}
 \newcommand{\bb}{\begin{thebibliography}}
 \newcommand{\eb}{\end{thebibliography}}

 \title{Intrinsic charm and the $D^+-D^-$ asymmetry produced in proton-proton
   collisions}

\author{G.I.~Lykasov$^1$, M.N.~Sorokovikov$^1$, S.J.~Brodsky$^2$}

\begin{document}
\maketitle
\begin{center}
  {\it $^{1}$Joint Institute for Nuclear Research, 141980, Dubna,
    Moscow region, Russia}\\
  {\it $^2$SLAC National Accelerator Laboratory, Stanford University,
    Menlo Park, CA 94025, United States}\\
\end{center}

\vspace{0.5cm}


{\bf Abstract }

We investigate the contribution    
of the charm-anticharm ($c{\bar c}$) asymmetry of the proton eigenstate 
obtained from QCD lattice gauge  
to the asymmetry of $D^+,D ^-$ and $D^0,{\bar D}^0$ mesons produced
in $pp$ collisions at large Feynman variables $x$.
It is shown that an important tool for establishing the intrinsic charm
({\it IC}) content
of the proton is the charm hadron-antihadron asymmetry formed in $pp$
collisions.
Predictions for the asymmetry as function of $x$
for different {\it IC} probabilities are presented.
We show that the interference of the intrinsic $|uud c{\bar c}>$ Fock state
with the standard
contribution from the PQCD evolution leads to a large $D^+D^-$ asymmetry at
large Feynman $x$. 

 
\section{Introduction} \indent

According to Quantum Chromodynamics (QCD), the heavy quarks in the nucleon sea
have both perturbative {\it extrinsic} and non-perturbative intrinsic heavy quark
({\it IQ})
components.   
The extrinsic sea quarks arise from gluon splitting, which is triggered by a probe in
the reaction. It can be calculated in QCD within perturbation theory.
In contrast,  the intrinsic sea quarks are contributions within the non-perturbative 
wave functions of the nucleon eigenstate which do not depend on the probe. 
The
existence of non-perturbative intrinsic charm {\it IC} was originally proposed in the
BHPS model~\cite{Brodsky:1980pb} and developed further in subsequent
papers~\cite{Brodsky:1984nx,Harris:1995jx,Franz:2000ee}.

In  Light-Front (LF) Hamiltonian theory, the intrinsic heavy quarks of the
proton are associated with non-valence Fock states, such as the $|uudQ{\bar Q}>$ Fock state in the 
hadronic proton eigenstate of the LF Hamiltonian; this implies that the heavy quarks
are multi-connected to the valence quarks, and thus have a probability which scales as $1\over M^2_Q$ or faster.
Since the LF wave function is maximal
at minimum off-shell invariant mass; i.e., at equal rapidity, the intrinsic
heavy quarks carry large light-front momentum fraction $x_Q$ in the hadronic eigenstate.

The concept of intrinsic heavy
quarks has also been proposed in the context of meson-baryon fluctuations~\cite{Navarra:1995rq,Pumplin:2005yf},
where intrinsic charm is
identified with the two-body hadron state $\bar{D}^0(u\bar{c})\Lambda^{+}_{c}(udc)$
in the proton.  Since these heavy quark
distributions depend on the nonperturbative correlations within the hadronic LF eigenfunction,
they also represent {\it intrinsic} contributions to the
hadron's fundamental structure.

A key characteristic of the non-valence LF wave functions
is the distinctly different momentum and spin distributions for the intrinsic $Q$ and anti-quark
${\bar Q}$ in the nucleon; for example, in the case of the charm-anticharm asymmetry,
the comoving heavy sea quarks are sensitive to the global quantum numbers of the
nucleon~\cite{Brodsky:2015fna}. Furthermore, since all of the intrinsic quarks
in the $|uudQ{\bar Q}>$ Fock state have similar rapidities, they can re-interact,
again, leading
to significant $Q$ vs ${\bar Q}$ asymmetries; i.e., large asymmetries in the charm
versus charm momentum and spin distributions.
The heavy quark asymmetry can also arise from the interference 
of the amplitudes corresponding to the intrinsic and extrinsic contributions.

Recently a new prediction for the non-perturbative intrinsic charm-anticharm
asymmetry of
the proton eigenstate has been obtained from a QCD lattice gauge
theory calculation of the proton's $G_{E}^p(Q^2)$ form
factor~\cite{Sufian:2020coz}. This form factor arises from the fact that the non-valence
quarks and anti-quarks have different distributions in the proton's
eigenstate. This result, together with the exclusive-inclusive connection
and analytic constraints on the form of hadronic structure functions from
Light-Front Holographic QCD (LFHQCD)~\cite{deTeramond:2005su}, predicts a significant non-perturbative
$c(x,Q) - \bar{c}(x,Q)$ asymmetry in the proton structure function at high
$x$, consistent with the dynamical features predicted by intrinsic charm models.

A recent measurement of the $c{\bar c}$ asymmetry 
has been reported by the NNPDF collaboration~\cite{NNPDF_Nov.2023}. The nonzero asymmetry
between the $D$ and ${\bar D}$ mesons extracted from  $Z+c$ production observed by the  LHCb
experiment in $pp$ collision~\cite{LHCb:2022} and at
EIC experiment~\cite{EIC:2021} in $eA$ collisions can be attributed
to the {\it IC} contribution in the nucleon PDF. There are also interesting
papers on the {\it IC} contribution to the proton PDF
\cite{Nadolsky:2023,Nadolsky:2018,Nadolsky2:2023} and \cite{Harland:2024}. 
  
  Despite intensive studies, the hypothesis
  of the non-zero intrinsic heavy quark contribution in the proton PDF
   \cite{Brodsky:1980pb} has not yet been definitively confirmed by experiment.
   There have been many attempts to confirm or estimate the {\it IQ} probability in the proton
   from experimental data; see \cite{BLLS:2020} and references therein.
   For example, the latest global analysis \cite{NNPDF:2022} shows that there is
  local evidence of the {\it IC} in the interval $0.3<x<0.6$ with confidence
   level CL of order 3$\sigma$. According to this analysis, the average fraction
   of the intrinsic charm quark is about $<x_c>=0.62\pm 0.28 \%$, including the
   PDF uncertainties.  However, the inclusion of missing high order
   uncertainties (MHOU) results in $<x_c>=0.62\pm 0.61 \%$, which is due to
   the large contribution of MHOU at $x<0.2$.  The reason for the large MHOU
   is related
   to the inclusion both 4FNS (four flavor number scheme), NNLO charm PDF determined from data, and 3FNS (three flavor number scheme,
   intrinsic) charm PDF within NNNLO corrections. Thus the determination of
   the {\it IC} content in the proton by analyzing only inclusive spectra of heavy hadrons or $c$-jets produced in $pp$
   collisions at LHC energies has proved to be very complicated.  Apart from these uncertainties, 
   there are statistical and systematical errors \cite{BBLMST:2019}.  
   In other words, various corrections to the LO approximation of the
   inclusive spectra calculations can reduce the evidence for {\it IC} contribution.  However, as we emphasize here, 
   the heavy hadron production
   asymmetries, for example, $D^+~D^-$ mesons produced in $pp$ collisions, are 
   sensitive to the {\it IC} contribution since the  nonperturbative $c{\bar c}$ pair is asymmetric.
   
   In this paper we analyze the heavy hadron asymmetry arising from nonperturbative sources of intrinsic charm in detail. For the calculation of the
   inclusive
   $D$-meson spectra and the $D~{\bar D}$ asymmetry in $pp$ collisions, we use
   the quark-gluon string model (QGSM). A short review of this approach and the scheme-dependence
   of its predictions is presented in Sect.2.
   We then show why the asymmetry of $D~{\bar D}$ mesons can be an ideal tool
 for identifying the {\it IC} content in the proton in comparison to the inclusive
   $D$-meson inclusive spectra. 

\section{Charmed hadron production in $pp$ collisions within QGSM without IC}   

The quark-gluon string model (QGSM) is based on
the $1/N$ topological expansion of
the amplitude of hadron-hadron interaction in the $s$-channel, which is related to its
$t$-channel expansion over Regge poles; here $N$ is the number of flavors
or colors \cite{t'Hooft:1974,Venez:1974,Kaidalov:1982,Capella:1994}.
This theory allows one to describe the data on hadron production in $pp$ collisions at
high energies and large $x$ rather satisfactorily
(\cite{Kaidalov:1980bq}-\cite{Arakelyan95}). 
The charmed hadron production in $pp$ collisions was considered recently within the
QGSM in \cite{Sinegovsky_2020}, where it was
 applied to the analysis of the 
 atmospheric neutrino flux. The satisfactory description of NA27 data
 \cite{NA27:1988}
for both charmed mesons and charmed baryons produced in $pp$ collisions at the
initial proton energy $\sim  400$  GeV was demonstrated. In ref.~\cite{Sinegovsky_2020}
the {\it IC} content in the proton was not included because the NA27 data were
measured at $x\leq 0.4$, where the {\it IC} contribution plays a minor role.
We will apply the QGSM to estimate the asymmetry of $D, {\bar D}$
mesons produced in $pp$ collisions, including the {\it IC} content
in the proton.           

The inclusive spectrum of hadrons produced in  $pp$ collisions as a function of the Feynman variable $x$ within
the QGSM is presented in the following form \cite{KaidalovPiskunova_ZPhys86,Sinegovsky_2020}:
\be
\label{equation:inclusiveCS}
\rho_h(x)=\int E\frac{d^3\sigma}{d^3p}d^2p_{\perp}=\sum\limits_{n=0}^{\infty}\sigma_{n}(s)\varphi_{n}^{h}(s,x),
\label{def:rho_h}
\ee
\noindent
where  $\sigma_n(s)$ is  the  cross  section  of  $2n$-strings (chains)
production, corresponding   to  the  
$s$-channel discontinuity of the multi-pomeron diagrams. The analysis
is performed for  
pomerons with $n$ cuts and an arbitrary number of external pomerons taking into
account elastic re-scattering.
Here $\varphi_{n}^{h}(s,x)$ is the $x$-distribution of the hadron $h$ produced  in
the fission  of $2n$ quark-gluon  
strings: $\varphi_{0}^{h}(s,x)$ accounts for the contribution of the 
diffraction  dissociation of colliding  hadrons, $n=1$ corresponds to the
strings formed  by valence  quarks  and 
diquarks; terms with  $n > 1$  are  related to  sea quarks  and antiquarks.

The cross sections $\sigma_{n}(s)$ were calculated ~\cite{Ter-Martirosyan} in
the quasi-eikonal approximation 
which accounts for the low-mass diffractive 
excitation of the colliding particles and corresponds 
to maximum inelastic diffraction consistent with the unitarity condition. Only
non-exchanged graphs were 
considered,  neglecting the interactions between pomerons.

In the case of $D$ meson production in $pp$ interaction, the functions
$\varphi_{n}^{h}(s,x)$  can be 
written~\cite{KaidalovPiskunova_charm} as follows:
\be
\varphi_{n}^{D}(s,x)= a^D\Big\{F_{q_V}^{D(n)}(x_{+})F_{qq}^{D(n)}(x_{-}) + F_{qq}^{D(n)}(x_{+})F_{q_V}^{D(n)}(x_{-})  \\  
+ 2(n-1)F_{q_{\rm sea}}^{D(n)}(x_{+})F_{{\bar q}_{\rm sea}}^{D(n)}(x_{-})\Big\},
\label{equation:phip}
\ee
where $x_{\pm}(s)=\frac{1}{2}\left[\sqrt{x^2+4m_{\perp}^{2}/s} \pm x\right]$.

The functions $F_{q_{V}}^{D(n)}(x_\pm)$, $F_{qq}^{D(n)}(x_\pm)$, and
$F_{q_{\rm sea}}^{D(n)}(x_\pm),F_{{\bar q}_{\rm sea}}^{D(n)}(x_\pm)$ are
defined as the convolution of the quark 
distributions with the fragmentation functions, taking into account
contributions of the valence quarks, diquarks, 
and sea quarks, antiquarks. For example, 
in $pp$ collisions~\cite{KaidalovPiskunova_ZPhys86,Shabelski,Lykasov:1999}:
\be
F_{q_{V}}^{D(n)}(x_\pm) = \frac{2}{3}\int\limits_{x_\pm}^{1} f_{p}^{u_{V}(n)}(x_1) G_{u}^{D}(x_\pm/x_1) dx_1 + \\
\nonumber
 + \frac{1}{3}\int\limits_{x_\pm}^{1} f_{p}^{d_{V}(n)}(x_1) G_{d}^{D}(x_\pm/x_1) dx_1,
\label{def:Fqv}
\ee
\be
F_{qq}^{D(n)}(x_\pm) = \frac{2}{3}\int\limits_{x_\pm}^{1} f_{p}^{ud(n)}(x_1) G_{ud}^{D}(x_\pm/x_1) dx_1 + \\
\nonumber
+ \frac{1}{3}\int\limits_{x_\pm}^{1} f_{p}^{uu(n)}(x_1) G_{uu}^{D}(x_\pm/x_1) dx_1.
\label{def:Fqq}
\ee
There are similar forms for $F_{q_{\rm sea}}^{D(n)}(x_{+})$ and $F_{{\bar q}_{\rm sea}}^{D(n)}(x_{-}) $. 
At the limits $x\rightarrow 0$ and $x\rightarrow 1$, these functions are defined by
Regge asymptotics. At 
intermediate values of $x$, an interpolation is 
used~\cite{KaidalovPiskunova_charm,KaidalovPiskunova_ZPhys86,Shabelski}. In particular, 
\begin{equation}
f_{p}^{u_{V}(n)}(x) = C^{u_{V}}_n x^{-\alpha_{R}(0)} (1-x)^{\alpha_{R}(0)-2\alpha_{N}(0)+n-1},
\label{def:uv_n}
\end{equation}
\noindent
\begin{equation}
G_{d}^{D^{-}}(x/x_1) = G_{\bar{u}}^{D^{0}}(x/x_1) = 
(1-x/x_1)^{\lambda-\alpha_{\psi}(0)}[1+a_1(x/x_1)^2],
\end{equation}
and the $ud$ diquark has the following $x$-dependence:
\begin{equation}
f_{p}^{ud(n)}(x) = C^{ud}_n (1-x)^{-\alpha_{R}(0)} x^{\alpha_{R}(0)-2\alpha_{N}(0)+n-1},
\label{def:ud_n}
\end{equation}
\noindent
where \, $\alpha_{R}(0)=0.5$, \, $\alpha_{N}(0)=-0.5$, \,
$\alpha_{\psi}(0)=-2.2$, \ \, 
$\lambda=2<p_{\perp}^{2}>\alpha^{\prime}_{R}=0.5$. The coefficient
$C^{u_{V}}_n$ is 
determined by the normalization $\int\limits_{0}^{1} f_{p}^{u_{V}(n)}(x)
dx=1$. One can see from Eqs.~(\ref{def:uv_n},\ref{def:ud_n}) that the diquark
distribution increases as $(1-x)^{-\alpha_{R}(0)}$, and the valence quark distribution
decreases as $(1-x)^{\alpha_{R}(0)-2\alpha_{N}(0)+n-1}$, when $x$ grows.  
More details on the functions 
$\varphi_{n}^{h}(s,x)$, $f_{p}^{j}(x)$ and $G_{j}^{D}(x/x_1)$ can be found 
in ref.~\cite{KaidalovPiskunova_charm,KaidalovPiskunova_ZPhys86,Shabelski,
Lykasov:1999,Arakelyan95}.

We note that the sea quark distributions within the QGSM have the same
forms as the valence quarks;  however, they 
contribute to inclusive hadron spectra starting from two-pomeron 
exchange, i. e., at $n\geq 2$.  
 
The distribution of the sea charmed quark in the proton has the following form
within the QGSM \cite{Lykasov:1999} 
(Appendix B): 
\be
f_{c{\bar c}}^{(n)}~=~C_{c{\bar c}} \delta_{c{\bar c}}x^{a_c} (1-x)^{b_c^n}~,
\label{def:fcbarc}
\ee
 where $a_c=-\alpha_\psi(0)$ and
 $b_c^n=2\alpha_R(0)-2\alpha_N(0)-\alpha_\psi(0)+n-1$ are presented in \cite{Lykasov:1999}, TABLE VII;  
$\delta_{c{\bar c}}$ is the weight of the sea $c{\bar c}$ in the proton. The coefficient $C_{c{\bar c}}$ 
is determined from the normalization condition
\be
\int_0^1 f_{c{\bar c}}^{(n)}(x) dx~=~C_{c{\bar c}}
\label{def:norm_fcc}
\ee
\be
C_{c{\bar c}}=\frac{\Gamma(2+a_c+b^n_c)}{\Gamma(1+a_c)\Gamma(1+b^n_c)} .
\label{def:coeffCi}
\ee 
Here $\Gamma(\alpha)$ is the gamma function.
The distribution of the sea bottom quark in proton has the following form within QGSM \cite{Lykasov:1999}: 
\be
f_{b({\bar b})}^{(n)}~=~C_{b({\bar b})} \delta_{b({\bar b})}x^a_b (1-x)^{b_b^n}~,
\label{def:fbbarb}
\ee
 where $a_b=-\alpha_Y(0)$ and
 $b^n_b=2\alpha_R(0)-2\alpha_N(0)-\alpha_Y(0)+n-1$, $\delta_{b({\bar b})}$ is
 the weight of the sea $b{\bar b}$ pairs in the proton. The normalization
 coefficient $C_{b({\bar b})}$ has the similar form as Eq.~(\ref{def:coeffCi})
 by replacing $a_c,b^n_c$ into $a_b,b^n_b$.

 \section{Charmed hadron production in $pp$ collisions within the QGSM including IC}

 We now include the contribution of {\it IC} in the sea charmed quark
distributions obtained within the QGSM.  We modify
Eq.~(\ref{def:fcbarc}) to the form
\be
f_{c{\bar c}}^{(n)}(x)\rightarrow (1-w)f_{c{\bar c}}^{(n)}(x)+wf^{in}_{c{\bar c}}(x),
\label{def:IC}
\ee
where $w$ is the probability of the {\it IC} contribution in
the proton and
$f^{in}_{c{\bar c}}$ is the intrinsic 
charm contribution to the conventional
charm distribution $f_{c{\bar c}}^{(n)}(x)$ at $n\geq 2$ given by Eq.~(\ref{def:fcbarc}). The form of
$f^{in}_{c{\bar c}}$ in the 
case of a symmetric $c{\bar c}$ sea in the proton, calculated within the BHPS model \cite{Brodsky:1980pb,BPS_1981} at $Q^2=m_c^2$, has
the following 
form \cite{Bluemlein_2016}:
\be
f^{in}_{c{\bar c}}(x)~=~600x^2\left\{(1-x)(x^2+10x+1)+6x(x+1)ln(x)\right\}~,
\label{def:BHPS}
\ee 
with the normalization condition 
\be
\int_0^1 f^{in}_{c{\bar c}}(x)dx~=~1.
\label{def:BHPS_norm}
\ee
The asymmetric intrinsic $f^{in}_{c({\bar c})}(x)$ distributions obtained
within the meson cloud-model, where the $|{uudc}{\bar c}\rangle$ Fock state can be
identified with the $|\Lambda_{udc} D_\mathrm{u\bar{c}} \rangle$ 
off-shell excitation of the proton. This distribution is parameterized in the
following form \cite{Pumplin_2007}: 
\be
f^{in}_{c}(x)~=~B x^{1.897} (1-x)^{6.095}~
\label{def:c_intr}
\ee
and
\be
f^{in}_{\bar c}(x)~=~~{\bar B} x^{2.5} (1-x)^{4.929}~,
\label{def:cbar_intr}
\ee
where $B/{\bar B}$ is determined by the quark number sum rule :
\be
\int_0^1 \left\{{f^{in}_{c}(x)-f^{in}_{\bar c}(x)}\right\}dx~=0~.
\label{def:normNbN}
\ee

  Generally the probability of the intrinsic heavy quark-anti quark  $Q{\bar Q}$ contribution in the proton is proportional to
  $1/M_Q^2$. It means that the coefficients $B$ and ${\bar B}$ are proportional to $1/M_Q^2$.  
The light-front $x$-distribution $f^{in}_{\bar c}(x)$ for the intrinsic
   $\bar c$ quarks is calculated as follows:
\be
f^{in}_{\bar c}(x)=f^{in}_{c}(x)-\Delta c(x)~,
\label{def:barc_intr_SB}
\ee   
where the difference ${\Delta c(x)}=[c(x)-{\bar c}(x)]$, obtained using
lattice QCD, was taken from \cite{Sufian:2020coz}.

\begin{figure*}
  \includegraphics[width=40pc]{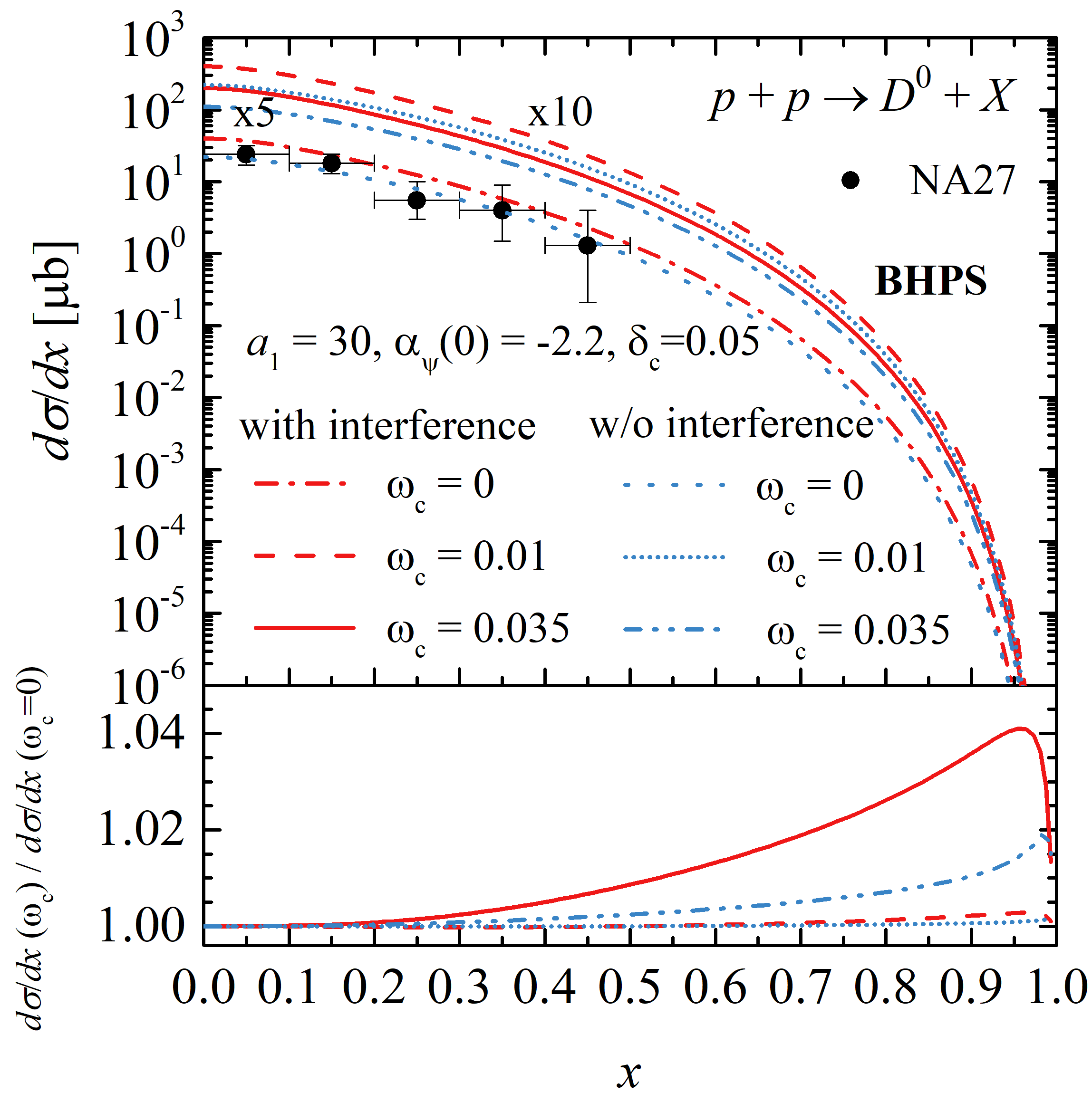}
	  \caption{\label{fig1}  The $x$-distribution of the inclusive spectrum 
        of $D^0$-mesons produced in $pp$ collisions at
        the initial energy $E_{lab}=$ 400 GeV and $0\leq x \leq 1$.
        Top: the inclusive spectrum at different values of $\omega$
        with and without (w/o) the interference of the quark (diquark)
        and the sea quark (extrinsic and intrinsic) amplitudes. The
        interference between
        the extrinsic and intrinsic quark contributions are also taken into
        account;  see Appendix. The NA27 data were taken from \cite{NA27:1988}. 
        Bottom: the ratio of the $x$-spectrum for the nonzero {\it IC}
        probability $\omega_c$ to the spectrum at $\omega_c=$ 0, with 
        interference and without it.
        With interference: slid red line at $\omega_c=$ 0.035, dashed red
        line at $\omega_c=$ 0.01. Without interference (w/o):
        double dashed blue line at $\omega_c=$ 0.035,
        dotted blue line at $\omega_c=$ 0.01.
        The calculations were done using the
        {\it IC} distributions in forms of Eqs.(\ref{def:c_intr},\ref{def:cbar_intr}),
        labeled as {\bf BHPS}.
        }
\end{figure*}

In Fig.~(\ref{fig1}) the inclusive $x$-spectrum of $D^0$
mesons produced in $pp$ collisions at the initial energy in the laboratory 
system $E_{lab}=$ 400 GeV ($\sqrt{s}=$ 27.42 GeV) is presented. One can observe
a good description of NA27 data for $D^0$-mesons. The results are presented at
different
values of the {\it IC} probabilities $\omega=$ 0.01 (BHPS1) and $\omega=$ 0.035 
(BHPS2), including both the important interference effect and without it; see
Appendix. One can see that the maximum signal of the {\it IC} contribution to
the spectrum is less than 1\%.
The NA27 data on $D^\pm$ meson \cite{NA27:1988} production have large
error bars, much larger than the data on $D^0({\bar D}^0)$;  therefore,
we compare our calculations with the data for $D^0({\bar D}^0)$ mesons.    

An unfortunate effect of the asymmetric {\it IC} contribution is invisible
in the $x$-spectrum presented in Fig.1. This is due to the fast decrease of the
intrinsic $c$-quark distribution, calculated within the meson cloud model,
according
to Eq.~(\ref{def:c_intr}) \cite{Pumplin_2007}, compared to the increasing
diquark
contribution at large $x$, as $1/(1-x)^{1/2}$, according to Eq.~\ref{def:ud_n}.
It leads to the dominant contribution of the diquark distribution to the
$x$-spectrum at large  $x$. Therefore, the distributions
of sea quarks give
a negligibly small contributions at large $x$.   
However, in contrast to that, the effect of the asymmetric {\it IC} can be visible in
the asymmetry of $D^-, D^+$ mesons produced in $pp$ collisions, as it will be
shown in the next section. 
The fragmentation functions of quarks and 
diquarks into $D^0,{\bar D}^0$ and $D^-,D^+$ mesons are presented in the
Appendix.

\section{Alternative approach for identifying the IC content in
  the proton within the QGSM}

There is another approach for investigation of the {\it IC} content in
the proton
within the QGSM, which was suggested in ref.~\cite{Shabelski:1995}. Intrinsic
quarks are 
valence-like quarks, according to \cite{Brodsky:1980pb}. That is why in
ref.~\cite{Shabelski:1995},  the intrinsic $c$-quark (anti quark)
$x$-distributions in the proton,
which are calculated within the QGSM, have the forms similar to those
of valence quark distributions. 

The sea $c({\bar c})$  and $b({\bar b})$ quark distributions within the QGSM
\cite{Kaidalov:1982,Capella:1994,Lykasov:1999}  are
presented in Eqs.~(\ref{def:fcbarc},\ref{def:fbbarb}).  At low $x$ the
$c$-quark 
distribution is proportional to $x^{-\alpha_\psi(0)}$, whereas the valence quark
distributions are proportional to $x^{-\alpha_R(0)}$; $\alpha_\psi(0)$ is the
intercept of the $\psi$-trajectory, and $\alpha_R(0)$  is the intercept of
the Reggeon
trajectory. These intercepts are very different: $\alpha_\psi(0)=$ 0 or
-2.2 and $\alpha_R(0)=$1/2. In \cite{Shabelski:1995} the sea $c({\bar c})$
quark distribution in the proton is parameterized similarly to
the valence quark distribution:
\be
f_{c\bar c}(x,n)~=~C_cx^{-\alpha_R(0)}(1-x)^{n+1}~; n>1
\label{def:Shab_uc}
\ee              
The sea quark distributions
$f_{u(\bar u)},f_{d(\bar d)}$ and $f_{s(\bar s)}$
in \cite{Shabelski:1995} are parameterized in the following forms:
\be
f_{u(\bar u)}^{sea}(x,n)=f_{d(\bar d)}^{sea}(x,n)=C_{\bar  u}x^{-\alpha_R(0)}
[(1+\delta/2)(1-x)^{n+\alpha_R(0)}(1-x/3)-\delta/2(1-x)^{n+1}]~,~n>1\\
\nonumber
f_{s}^{sea}(x,n)=C_sx^{-\alpha_R(0)}(1-x)^{n+1}~,~n>1~,
\label{def:Shab_fsea}
\ee
The convolutions of sea quarks with their fragmentation functions
to the $D$-meson,
see Eqs.~(\ref{def:Fqv},\ref{def:Fqq}), are given in the following forms:
\be
F_{q_{\rm sea}}^{D(n)}(x_{\pm})=\frac{1}{2+\delta_s+
  \delta_c}[F_{u_{\rm sea}}^{D(n)}+F_{d_{\rm sea}}^{D(n)} 
+\delta_sF_{s_{\rm sea}}^{D(n)}+\delta_cF_{q_{\rm sea}}^{D(n)}]
\label{def:conv_seaq}
\ee 
The fragmentation functions (FF) of $c,{\bar c}$ quarks to $D$ mesons were
taken from
\cite{Shabelski:1995} in the forms similar to the FF into $K$-mesons, namely:
\be
G_{(c+{\bar c})/2}^{D^+}=G_{(c+{\bar c})/2}^{D^0}=G_{(c+{\bar c})/2}^{{\bar D}^0}=G_{(c+{\bar c})/2}^{D^-}=
[a_{c0}z(1-z)^{\lambda-\alpha_R(0)}+a_{c1}(1-z)^{1+\lambda-\alpha_R(0)}]~, 
\label{def:conv_seaFF}
\ee  
where $a_{c0}=$0.68, $a_{c1}=$0.26; $\alpha_R(0)=1/2$; 
$\delta=\delta_s+\delta_c$. Here $\delta_s=$0.2 and $\delta_c=$0.04 are the relative
probabilities of strange and charmed quarks existing in the proton sea
\cite{Shabelski:1995}.

In order to compare our calculations within the QGSM with the similar
calculations done in \cite{Shabelski:1995}, we present the following ratios
$R_1,...R_4$:\\

$R_1=((\frac{d\sigma}{dx})_1$ at $\delta_c=$0.01,$\delta_s=$0.2)$/\sigma_s$\\

$R_2=((\frac{d\sigma}{dx})_2$ at $\delta_c=$0.04,$\delta_s=$0.2)$/\sigma_s$\\
 
$R_3=((\frac{d\sigma}{dx})_{S}$ at $\delta_c=$0.04,$\delta_s=$0.2)$/\sigma_s$\\

$R_4=((\frac{d\sigma}{dx})_{S}$ at $\delta_c=$0.01,$\delta_s=$0.2)$/\sigma_s$,\\

where $\sigma_s\equiv \frac{d\sigma}{dx}$ is calculated within the QGSM
in ref.~\cite{Shabelski:1995} using sea
quark distributions at $\delta_c=$0,$\delta_s=$0;
$(\frac{d\sigma}{dx})_{1,2}$ is the
$x$-spectrum of $D^0({\bar D}^0)$ calculated within the QGSM in this paper;
$(\frac{d\sigma}{dx})_{S}$ is the $x$-spectrum calculated in
\cite{Shabelski:1995}.
        \begin{figure*}
	\centering
	\begin{minipage}{17pc}
		\includegraphics[width=20pc]{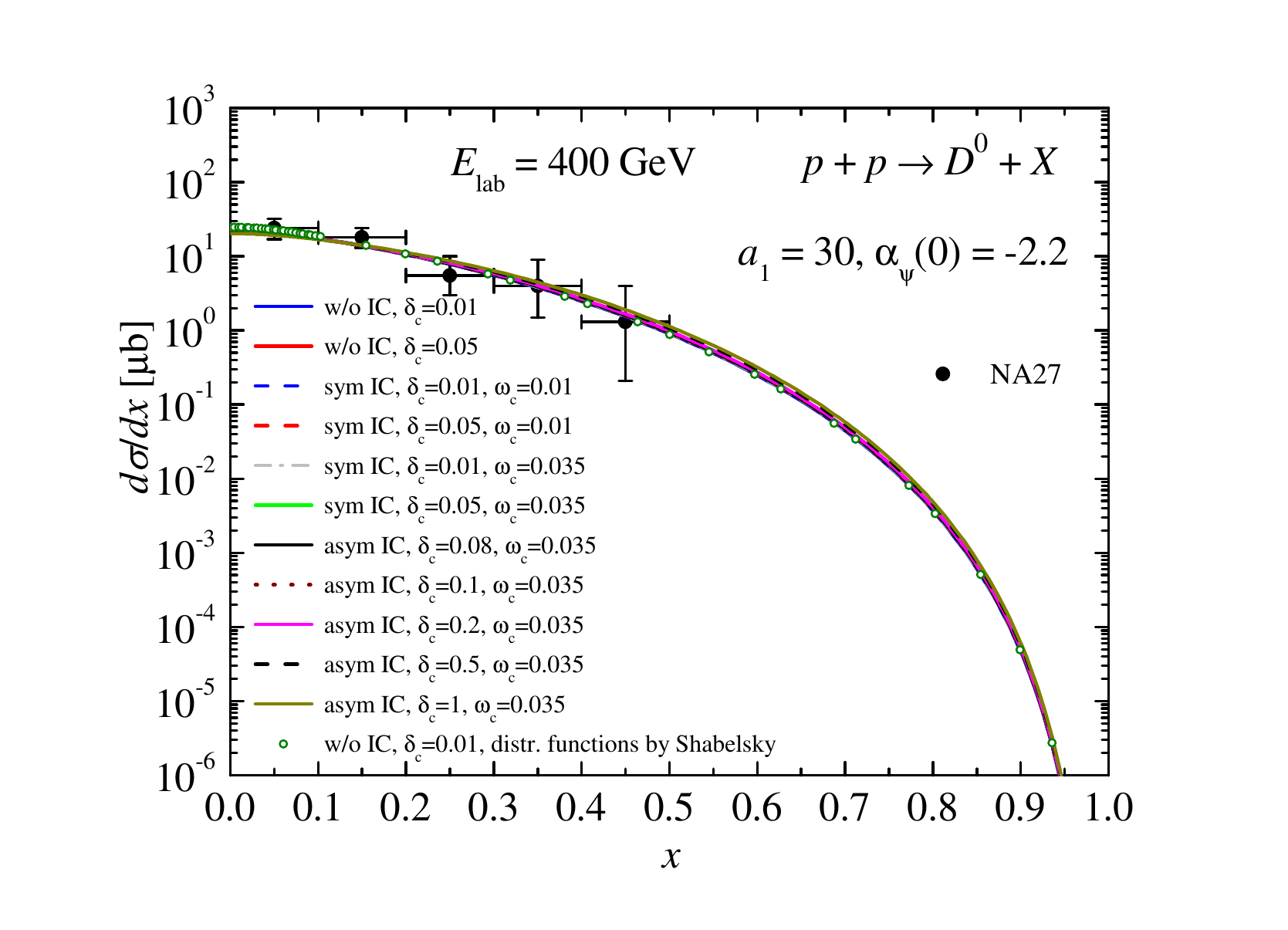}
	\end{minipage}
	\hspace{1pc}%
        \qquad
	\begin{minipage}{17pc}
	\includegraphics[width=20pc] {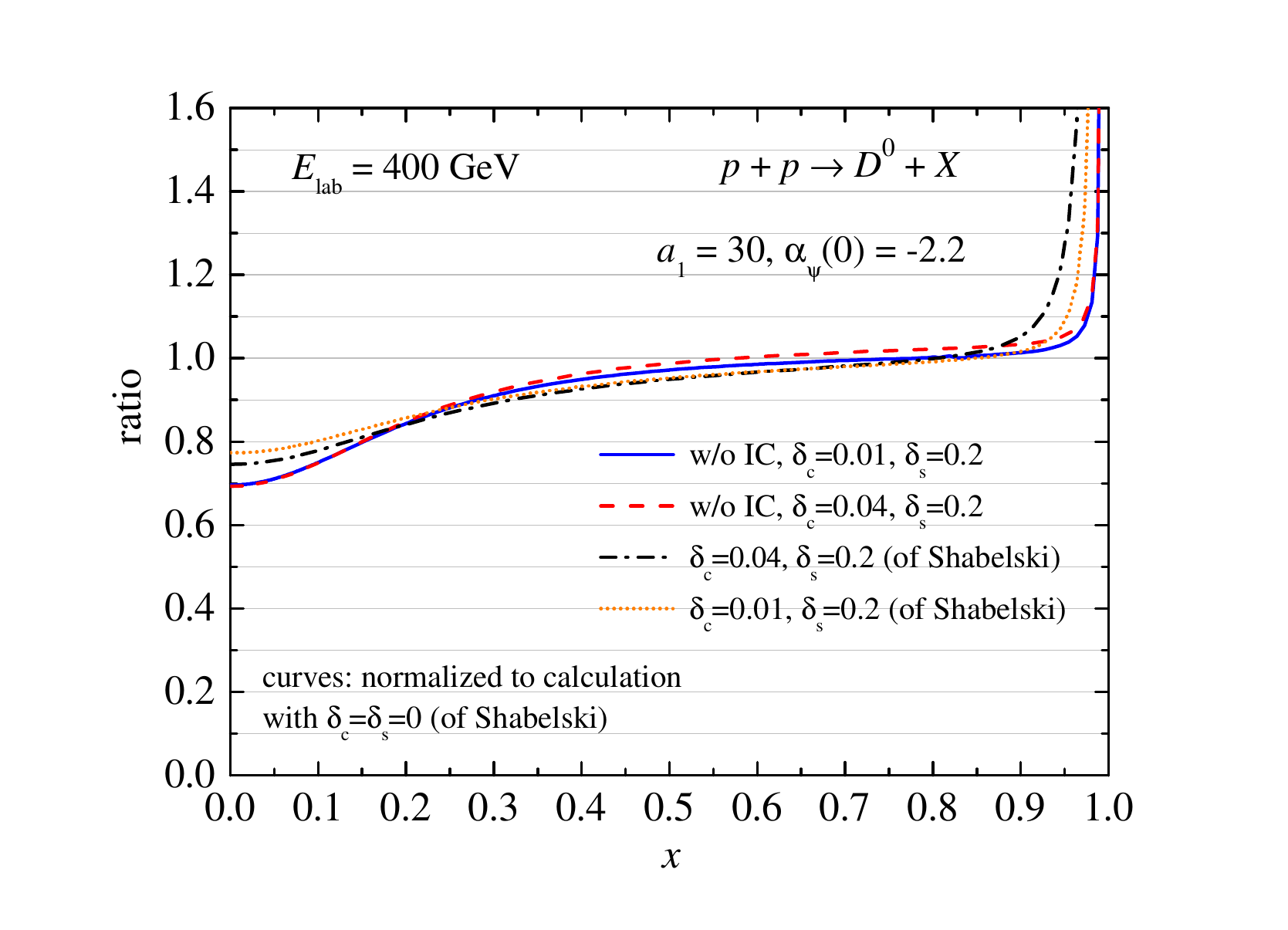}  
	\end{minipage} 
	\caption{\label{fig2} Left: our calculation of the inclusive
          $x$-spectrum of $D^0$-mesons produced in $pp$
          collisions at the initial energy $E_{lab}=$ 400 GeV  with different
          values of the weight $\delta_c$ and the {\it IC} probability $w_c$.
          The open circles correspond to the calculation of ref.~\cite{Shabelski:1995}.  
          Right: the blue solid line is the ratio $R_1$, the red dashed line is
          $R_2$, the black dash-dotted line is $R_3$ and the orange dotted line
          is $R_4$.       
            }
\end{figure*}
 In Fig.~(\ref{fig2}, left), the $x$-spectrum of $D^0$-mesons produced in $pp$
 collisions at $\sqrt{s}=$27 GeV is presented for different weights of the sea 
 $c{\bar c}$ and $s{\bar s}$ pairs and the {\it IC} weight $w_c$.
 One can see from
 this panel,  that its sensitivity to these parameters is very small. 
 In Fig.~(\ref{fig2}, right) the ratios $R1,..,R_4$ are presented as
 functions of $x$. The difference between
 our calculation within the QGSM from the one within the model of
 \cite{Shabelski:1995} is about 10\%-13\% at small $0\leq x\leq 0.15$.
 This is due to different forms of the sea quark distribution.
 In \cite{Shabelski:1995} it is assumed that both the
 extrinsic and intrinsic charm quarks have the same distributions as the valence
 quarks $f_p^{u_v^n}(x),f_p^{d_v^n}(x)$  at $n>$1, where $n$ is the number
 of exchanged Pomerons
 or $2n$ quark-antiquark pairs. In some sense, the charmed and strange
 quarks have valence-like distributions, as it was assumed for the intrinsic 
 $Q{\bar Q}$ pairs in \cite{Brodsky:1980pb}.
 In \cite{Shabelski:1995} the weight $\delta_c$ of sea $c{\bar c}$ pairs
 includes
 both extrinsic and intrinsic charmed quarks.        
 Comparing Fig.~(\ref{fig1}) with Fig.~(\ref{fig2}), the question arises why
 there
 is no visible difference between the two QGSM approaches presented in Sections
 (2,3) and 4, although the $c$-quark distributions in proton are different.
 This is due to the  main contribution of the diquark $x$-distribution at
 $x\rightarrow$ 1, according to Eq.(\ref{def:ud_n}), compared to the valence
 quark
 distributions;  see Eq.(\ref{def:uv_n}) at large $x$. Therefore, the inclusive
 $x$-spectra of
 $D,{\bar D}$-mesons are almost insensitive to the sea-quark distributions in the
 proton
 at large $x$. However, the asymmetry of $D^+,D^-$-mesons, as a function of
 $x$,
 can be visible at large $x$, as will be shown in the next section.  
 
 \section{ $D^0{\bar D}^0$ and $D^- D^+$ asymmetries in $pp$ collisions}

In an important recent development~\cite{Sufian:2020coz}, the difference of the nonperturbative
charm and anticharm quark distributions in the proton ${\Delta c(x)} = c(x) -
\bar{c}(x)$ has been computed using lattice
gauge theory. 
The predicted ${\Delta c(x)}$ 
distribution is non-zero at large $x \geq 0.4$, 
remarkably consistent with the
expectations of intrinsic charm. The $c(x)$ vs. $\bar{c}(x)$ asymmetry can be
understood physically by identifying the $|{uudc}{\bar c}\rangle$ Fock state
with the
$|\Lambda_{udc} D_\mathrm{u\bar{c}} \rangle$ off-shell excitation of the proton.

Let us calculate  the asymmetry $A_{D^0 {\bar D}^0}(x)$  and
$A_{D^- D^+}(x)$ for $D$-mesons produced in $pp$ collisions, 
including the asymmetry between the intrinsic 
$c$-quark and ${\bar c}$ quark as a function of $x$.  

\be
A_{D^0 {\bar D}^0}(x)~=~\frac{d\sigma_{D^0}/dx - d\sigma_{{\bar D}^0}/dx}
{d\sigma_{D^0}/dx + d\sigma_{{\bar D}^0}/dx}
\label{def:asymD0barD0}
\ee
\be
A_{D^- D^+}(x)~=~\frac{d\sigma_{D^-}/dx - d\sigma_{D^+}/dx}
{d\sigma_{D^-}/dx + d\sigma_{D^+}/dx}
\label{def:asymDplDmin}
\ee
The calculation of these asymmetries was done within two sets.
The first one {\it Set 1}
  corresponds to use Eqs.(\ref{def:c_intr},\ref{def:cbar_intr}) for distributions $f_c^{in}(x)$
  and $f_{\bar c}^{in}(x)$ respectively obtained in \cite{Pumplin_2007}  within the
  the meson cloudy bag model. {\it Set 2} was performed using Eq.\ref{def:c_intr}
  for $f_c^{in}(x)$
  and calculating $f^{in}_{\bar c}(x)$ from Eq.~(\ref{def:barc_intr_SB}) because
  only $\Delta c(x)$
  was calculated in \cite{Sufian:2020coz} within the lattice QCD.

\begin{figure*}
	\centering
	\begin{minipage}{17pc}
	  \includegraphics[width=20pc]{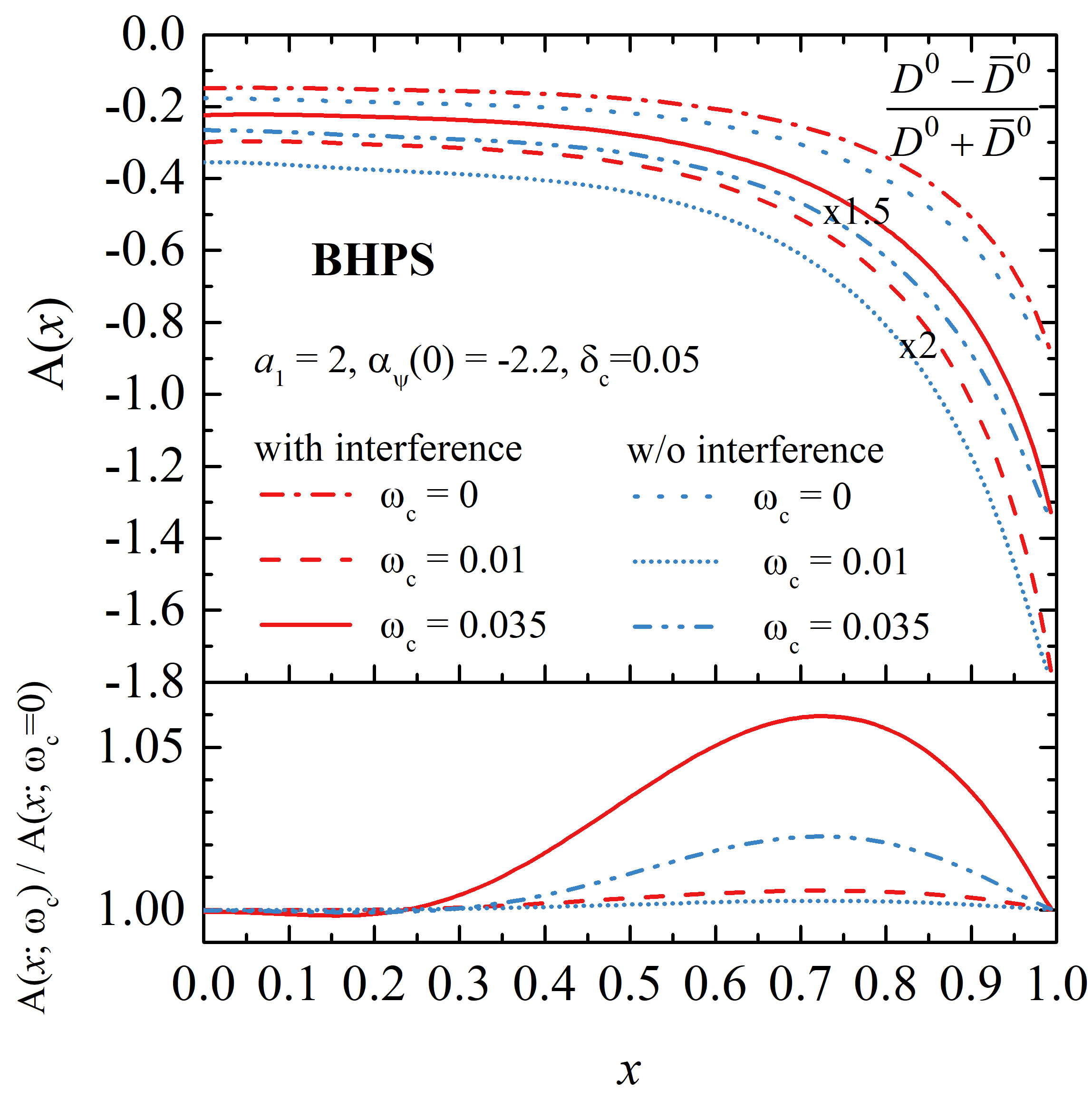}
	\end{minipage}
	\hspace{1pc}%
       \qquad
	\begin{minipage}{17pc}
	  \includegraphics[width=20pc]{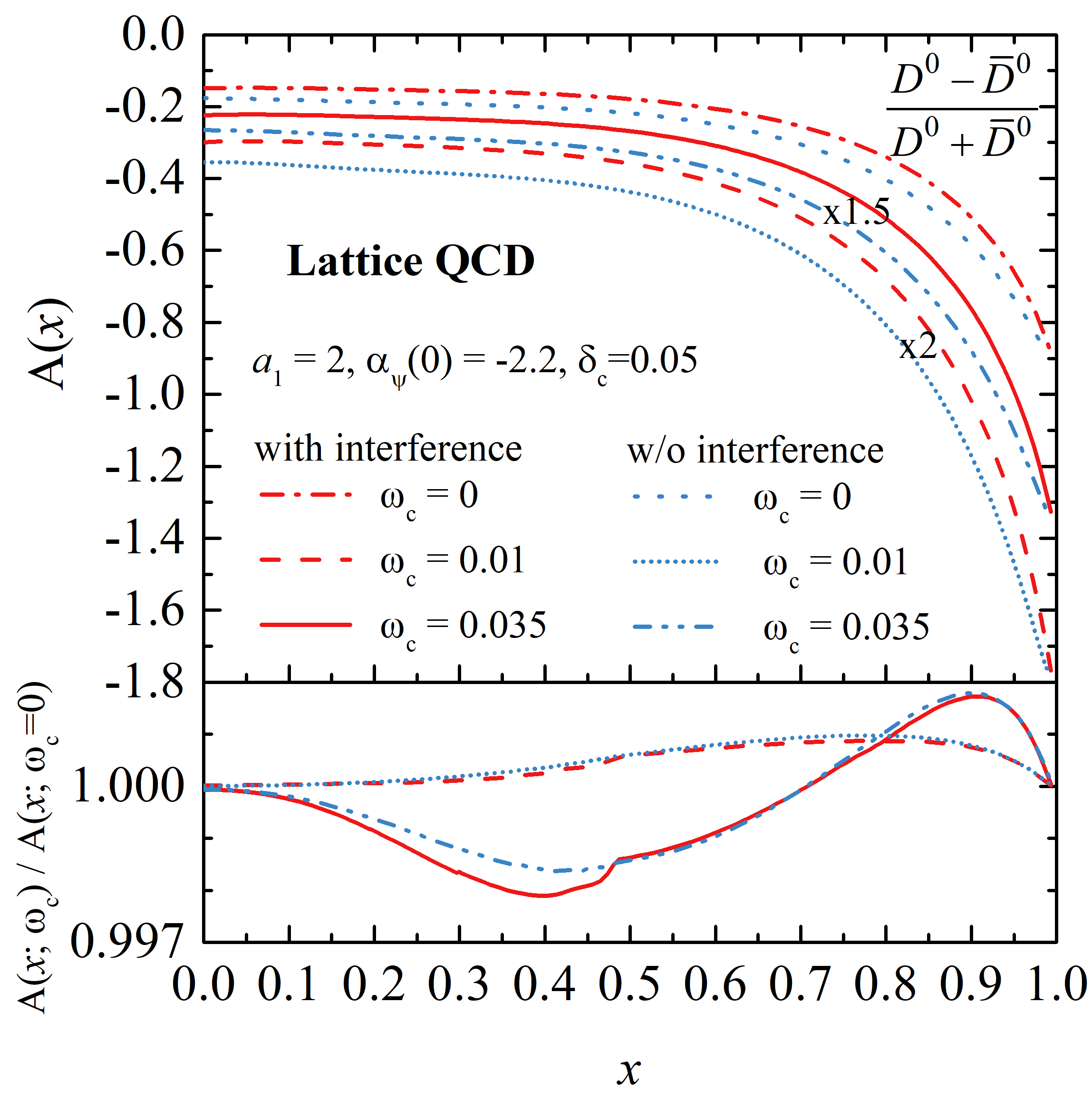}
          \end{minipage} 
	\caption{\label{fig3} Left:  the {\it Set 1} of $x$-distribution of asymmetry
        $A_{D^0{\bar D}^0}(x)$ 
        between $D^0$ and ${\bar D}^0$- mesons produced in $pp$ collision at
        the initial energy $E_{lab}=$ 400 GeV and $a_1=$ 2, and the ratio of
        the asymmetry with the non zero {\it IC} probability $\omega_c$ to
        the one with $\omega_c=$ 0. The calculations were done using the
        {\it IC} distributions in forms of Eqs.(\ref{def:c_intr},\ref{def:cbar_intr}).
        Right: the {\it Set 2} of asymmetry $A_{D^0{\bar D}^0}(x)$
        using Eq.(\ref{def:c_intr}) for $f_c^{in}(x)$
        and Eq.(\ref{def:cbar_intr}) for $f_{\bar c}^{in}(x)$. 
        With interference: solid red line at $\omega_c=$ 0.035, dashed red line
        at $\omega_c=$ 0.01. Without interference (w/o):
        double dashed blue line at $\omega_c=$ 0.035,
        dotted blue line at $\omega_c=$ 0.01.
            }
\end{figure*}


The results of $A_{D^0 {\bar D}^0}(x)$ with different {\it IC} probabilities
$\omega_c$ calculated within the QGSM are presented in Fig.~\ref{fig3}.
In the left panel the {\it Set 1} calculation is presented, whereas
the {\it Set 2} results are shown on the right.
The calculations were done taking into account the interference between
different parton contributions, also including the important interference between the
contributions of extrinsic and intrinsic quarks;  see
Eqs.(\ref{def:rho_D},\ref{def:rhotot},\ref{def:A1},\ref{def:A2}) in the Appendix. 

The inclusion of the interference
effects increases the asymmetry at $0< x < 0.4$ by a maximum of (25-30)\%
for both calculations {\it Set 1} and {\it Set 2}, as it is
seen in Fig.~\ref{fig3} (top). Both calculations show too small {\it IC} contribution
for the $D^0 {\bar D}^0$ asymmetry as it is seen from the bottom of this figure.

\begin{figure*}
	\centering
	\begin{minipage}{17pc}
	  \includegraphics[width=20pc]{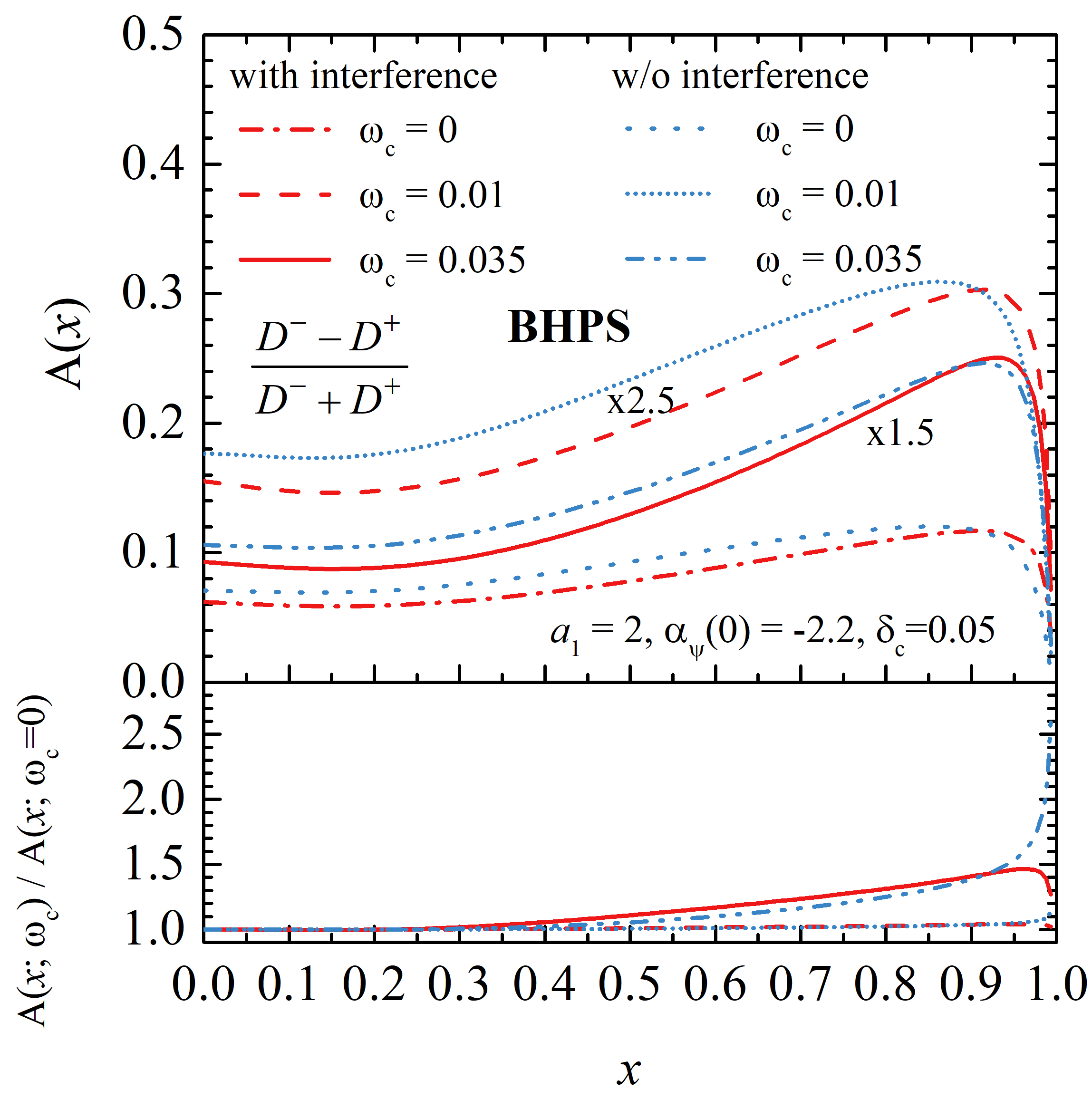}
	\end{minipage}
	\hspace{1pc}%
        \qquad
	\begin{minipage}{17pc}
	  \includegraphics[width=20pc]{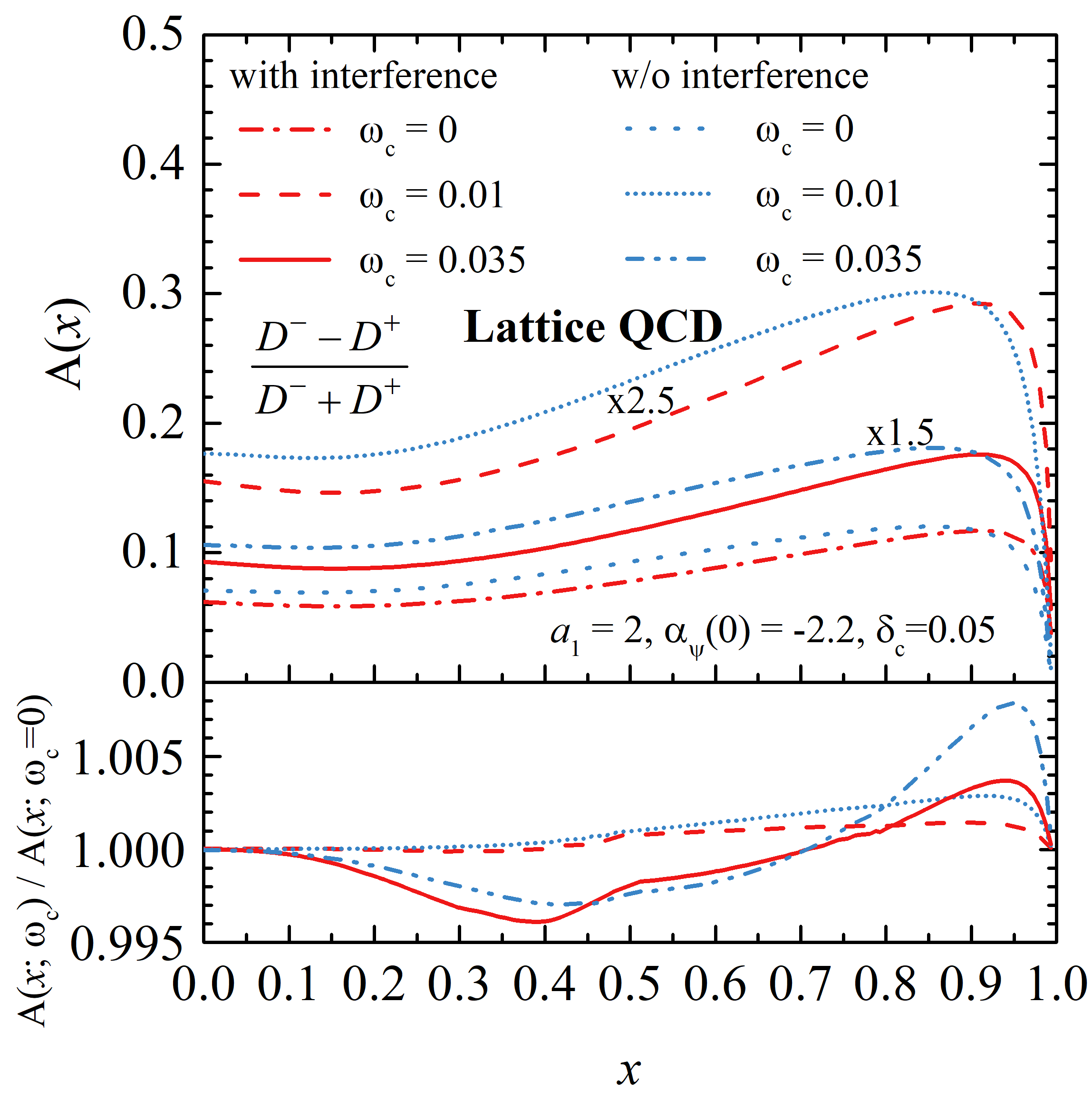}
          \end{minipage} 
	\caption{\label{fig4}
        The plots for asymmetry $A_{D^-D^+}(x)$ with the same notations as in Fig.\ref{fig3}.
        }
        \end{figure*}

The asymmetry $A_{D^-  D^+}(x)$ as a function of $x$
using {\it Set 1} (left) and {\it Set 2} (right) is presented in 
Fig.~\ref{fig4}.
In the bottom of these plots, the ratios of the asymmetry for non-zero {\it IC}
probability $\omega_c$ to the asymmetry at $\omega_c=$ 0 are presented.
The {\it Set 1} is more sensitive to the {\it IC} probability
$\omega_c$ compared to {\it Set 2}, if we compare the bottom plots in Fig.~\ref{fig3}.
The difference between the left and right panels of Figs.~(\ref {fig3},\ref {fig4})
is due to the absence of information on the separate distributions $c(x)$ and ${\bar c}(x)$
obtained within the {\bf Lattice} QCD, in  \cite{Sufian:2020coz} only $\Delta c(x)$ is presented.
It is not sufficient to calculate the asymmetry precisely.
The maximum value of the $D^ - D^+$
asymmetry could be about 50\% at $x\simeq$ 0.9 at $a_1=$ 2
(right bottom in figure). Both bottom panels of Fig.~\ref{fig4} show
a sizable sensitivity of the asymmetry to the interference effects at
large $x$. The interference effects for the $D^-D^+$ asymmetry increase, 
when $x$ grows, as Fig.~\ref{fig4} shows. 
Note that the interference
between different amplitudes estimated in Appendix is maximal,
when the relative phase angle between different amplitudes is approximately zero,
i.e., the constructive sum of the amplitudes is calculated.   
The form of the $A_{D^-  D^+}(x)$ enhancement at large $x$ corresponding to the
positive difference $d\sigma_{D^-}/dx - d\sigma_{D^+}/dx$ at 0.5$<x<$1.0 is
determined mainly by the FF of quarks and diquarks to
$D^-$ or $D^+$, which describe the inclusive $x$-spectra of the open charm
 within the QGSM rather satisfactorily 
\cite{Kaidalov:1982}-\cite{Arakelyan95}. The FF used in our calculations
of $x$-spectra and the asymmetries of $D,{\bar D}$-mesons are presented
in the Appendix.      


        \begin{figure*}
	\centering
	\begin{minipage}{17pc}
	  \includegraphics[width=20pc]{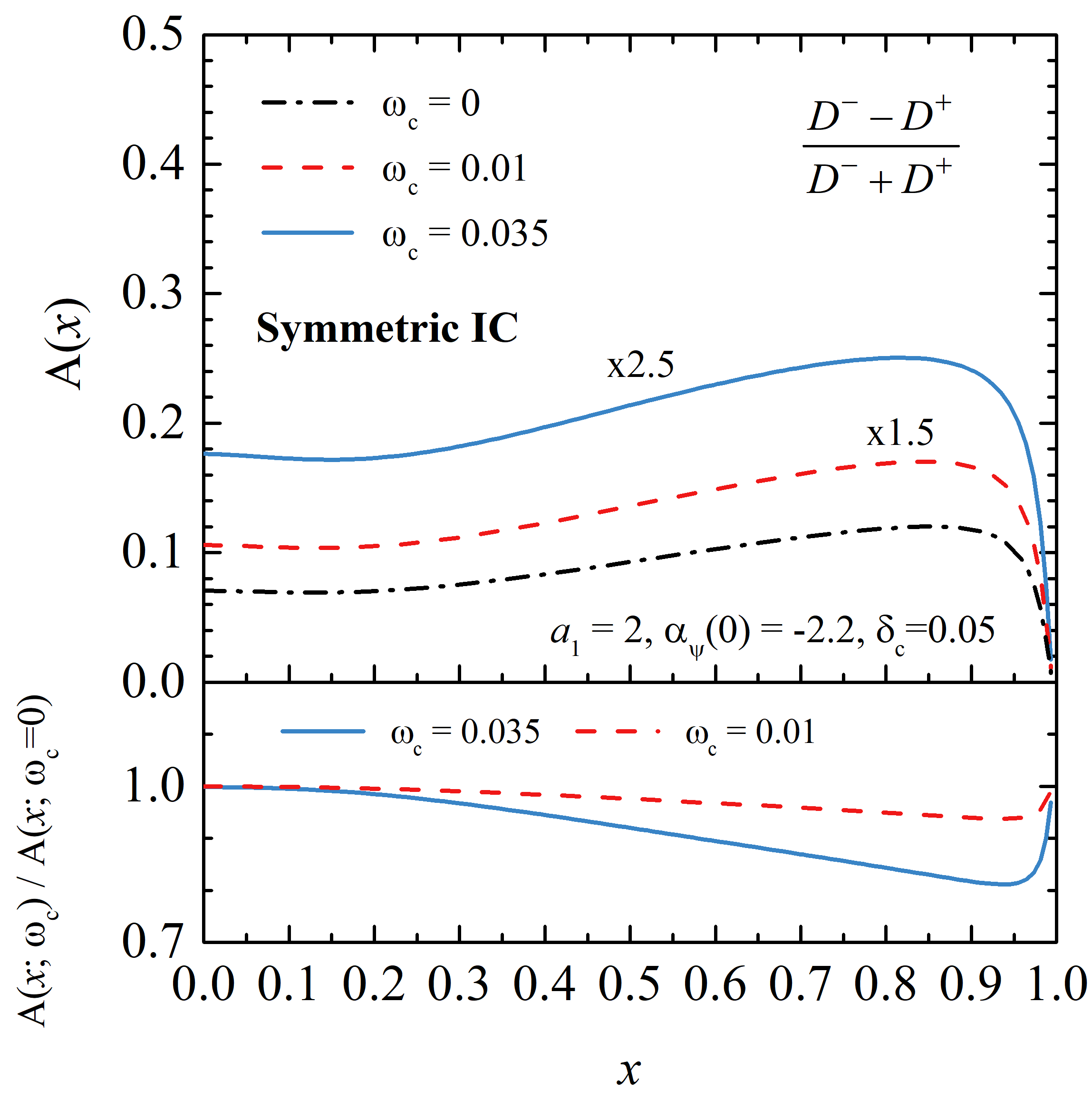}
	\end{minipage}
	\hspace{1pc}%
        \qquad
	\begin{minipage}{17pc}
	  \includegraphics[width=20pc]{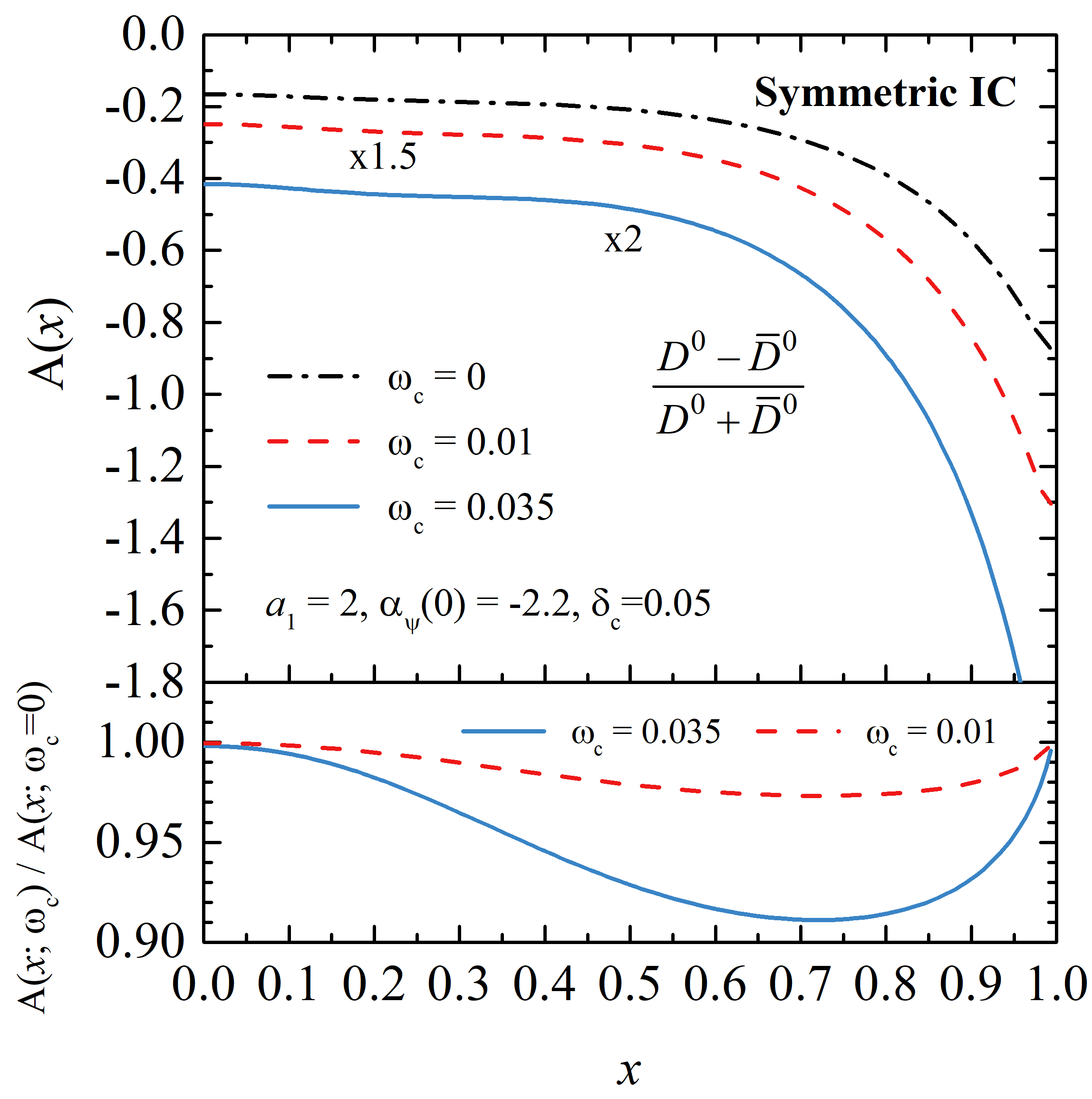}
          \end{minipage} 
	\caption{\label{fig5}
          The  plots of the asymmetries $A_{D^-D^+}(x)$ (left)
          and $A_{D^0{\bar D}^0}(x)$ (right) using the symmetric {\it IC} distribution,
          according to Eq.~(\ref{def:IC})
        with the same notations as in Fig.~\ref{fig3}.
        }
        \end{figure*}


         \begin{figure*}
	\centering
	\begin{minipage}{17pc}
	  \includegraphics[width=20pc]{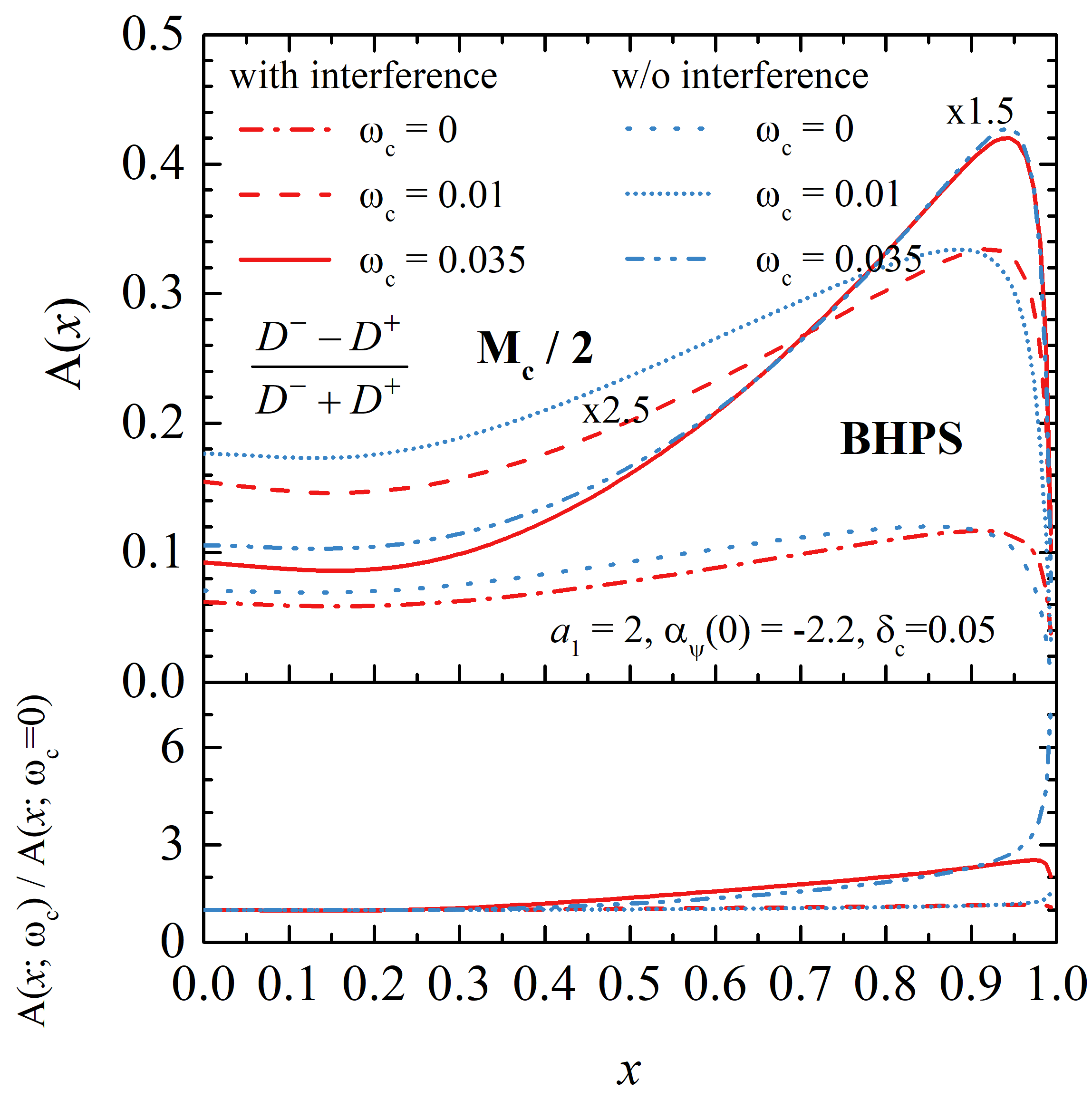}

	\end{minipage}
	\hspace{1pc}%
        \qquad
	\begin{minipage}{17pc}
	  \includegraphics[width=20pc]{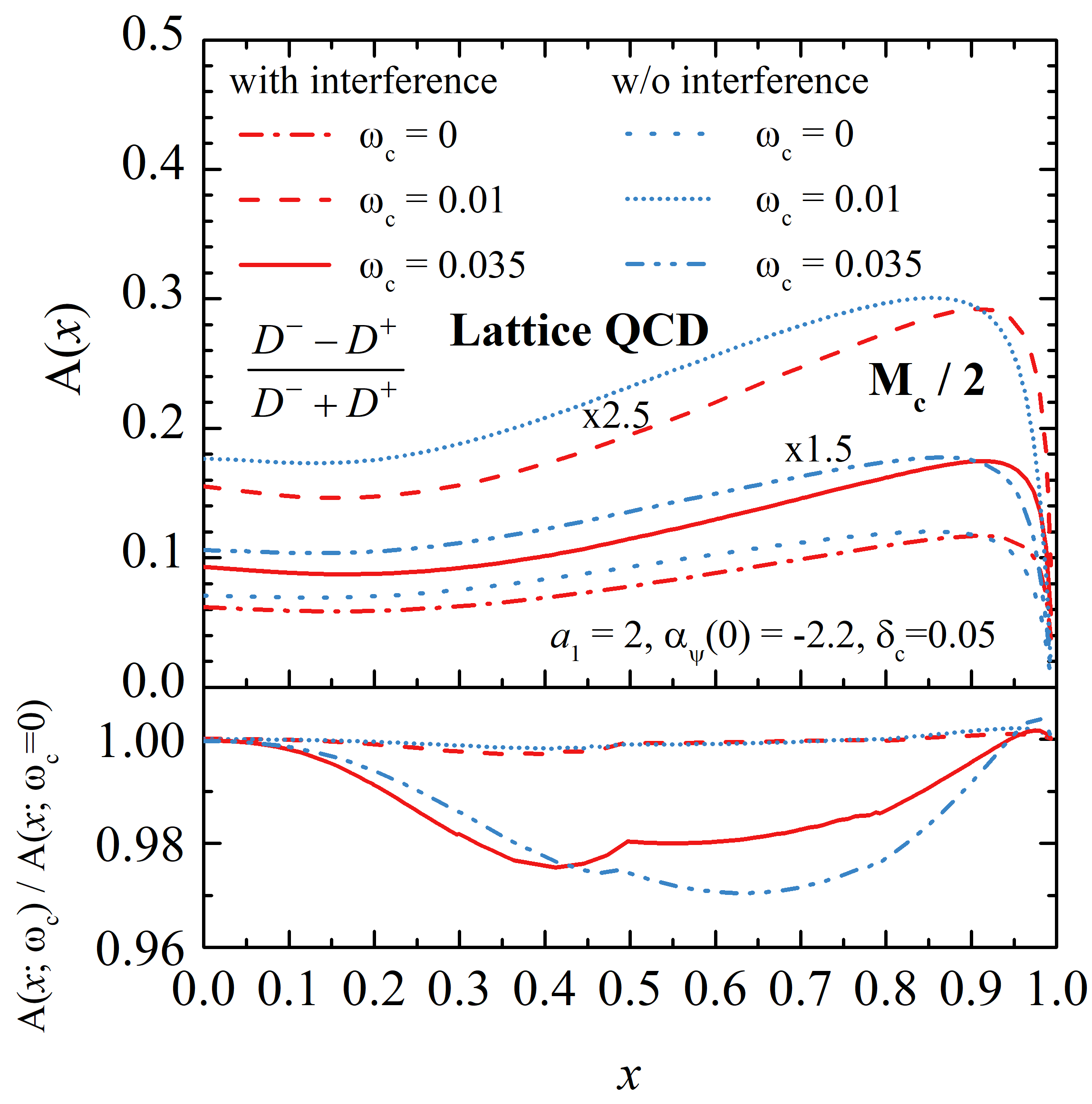}
          \end{minipage} 
	\caption{\label{fig6}
          The  plots of the asymmetry $A_{D^-D^+}(x)$ for the mass of intrinsic quark $M_Q=M_c/2$
          with the same notations as in Fig.~\ref{fig3}.
            }
         \end{figure*}

                  \begin{figure*}
	\centering
	\begin{minipage}{17pc}
	  \includegraphics[width=20pc]{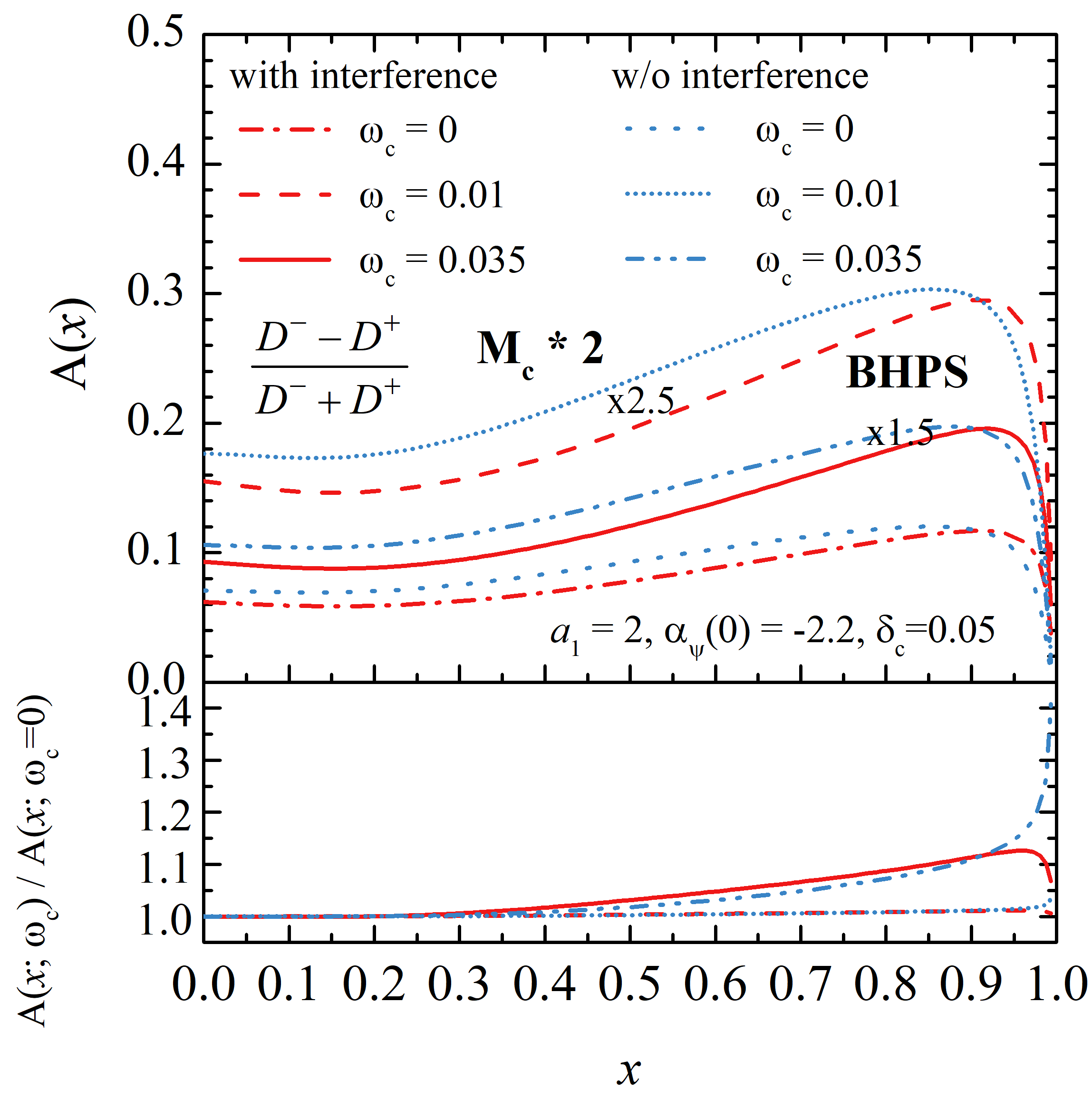}
	\end{minipage}
	\hspace{1pc}%
        \qquad
	\begin{minipage}{17pc}
	  \includegraphics[width=20pc]{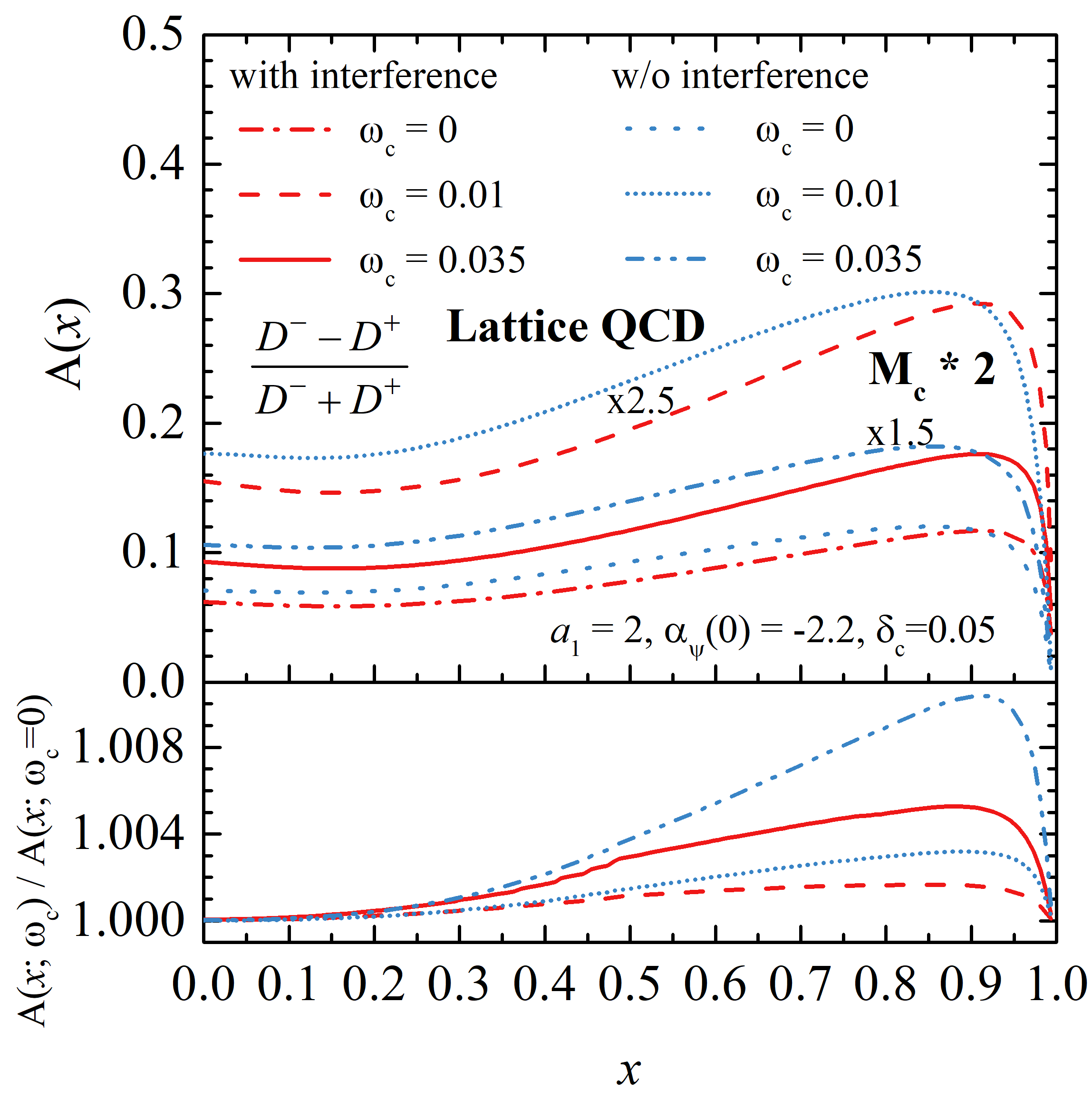}
          \end{minipage} 
	\caption{\label{fig7}
        The  plots of the asymmetry $A_{D^-D^+}(x)$ for the mass of intrinsic quark $M_Q=M_c*2$
        with the same notations as in Fig.~\ref{fig3}.
        }
                  \end{figure*}

We note, that for the calculation of the asymmetries presented in
Figs.~(\ref{fig3},\ref{fig4})
within the QGSM we used 
the charm (anti-charm) quark $x$-distributions $c(x)$ and ${\bar c}(x)$;
therefore $\Delta c(x)=c(x)-{\bar c}(x)$ and asymmetries
$A_{D^0 {\bar D}^0}(x)$, $A_{D^- {\bar D}^+}(x)$ are shifted
towards larger $x$ in comparison to $x\Delta c(x)$, calculated within the
lattice QCD in \cite{Sufian:2020coz}, and in the NNPDF paper~\cite{NNPDF_Nov.2023}.

The experimental data of WA82 \cite {WA82:1993} and E769 \cite {E769:1994} for 
$D{\bar D}$ production in $\pi^- p$ production at initial momenta 250 GeV$/$c
and 340 GeV$/$c  display an increase of the $D^-D^+$ asymmetry as a function
of $x$, consistent with the calculation of $A_{D^- D^+}(x)$ presented in
Fig.~\ref{fig4}.

  In Fig.(\ref{fig5}) the asymmetries $D^+D^-$ (left) and $D^0{\bar D}^0$ (right)
  are  presented for the symmetric {\it IC} distribution given by
Eq.~(\ref{def:IC}). Comparing Fig.(\ref{fig4}) and Fig.(\ref{fig5}) one can
see that both asymmetries change sign if the asymmetric {\it IC} distributions
are included. This is a very important signal for identifying the {\it IC}
contribution in the proton.
  
  
Let us study the sensitivities of $D^+ D^-$ asymmetry to the mass of intrinsic
heavy quarks $M_Q$. In Figs.(\ref{fig6},\ref{fig7})
the asymmetries as functions of $x$ are presented at $M_Q=M_c/2$ and
$M_Q=M_c*2$. One can see from these figures that the {\it Set 1}
calculation, taking into account the {\it IC} contribution, shows the
increase of the asymmetry if $M_Q$ decreases.

The string model (QGSM) we have used, as it mentioned in
Section 2, is based on the $1/N$ topological expansion of
the $h-h$ interaction amplitude proposed in refs.~ \cite{t'Hooft:1974,Venez:1974}
instead of an $\alpha_s$ expansion in perturbative QCD.
This approach, developed later in
ref.~\cite{Kaidalov:1982,Capella:1994} operates with the quark (valence and sea)
and diquark distributions in the proton, and gluons are considered as
$q{\bar q}$. Subprocesses of type the gluon splitting
into $c{\bar c}$, which result in the {\it extrinsic } $c{\bar c}$ pairs,
as calculated within perturbative QCD, are considered as the sea
charm-anticharm pairs, taking into account their weight $\delta_c$
(Eq.~\ref{def:fcbarc})
determined from a fit of experimental data for the open charm production.
A central question is the contributions of {\it extrinsic } and
{\it intrinsic} $c{\bar c}$ pairs and their interference to
the $D^- D^+$ asymmetry. We have estimated it within the QGSM, and the results are
presented in Fig.~\ref{fig4}.
  
\section{Conclusion}

The $D$-meson production in $pp$ collisions predicted by nonperturbative QCD,
taking into account
the intrinsic charm content in the proton has been analyzed.
We have investigated the relation of the $c{\bar c}$ non-perturbative intrinsic
charm-anticharm asymmetry of the proton eigenstate as
obtained from a QCD lattice gauge theory calculation of the proton's
$G_{E}^p(Q^2)$ form
factor~\cite{Sufian:2020coz} to the asymmetry
$A_{D^- D^+}(x)$ and $A_{D^0 {\bar D}^0}(x)$ of $D$-meson
and ${\bar D}$-mesons produced in $pp$ collisions at
large $x$.

We have shown that the inclusion of the asymmetric {\it IC}
components in the proton as calculated within the meson cloudy-bag model
~\cite{Pumplin_2007} leads to the enhancement of the
$D^-D^+$ asymmetry at $x>$0.7. The {\it IC} signal of this asymmetry
depends on the intrinsic charm probability $w_{c}$.  The maximal deviation of
the $D^-D^+$ asymmetry compared to the prediction for zero {\it IC} is about 50\% at
$x \simeq$ 0.9 and $w_{IC}=$3.5\%.
In principle, such a deviation in the range of
15\%-20\% can be experimentally observed.
   
As for the  $D^0 {\bar D^0}$ asymmetry in $pp$ collisions,
the {\it IC} signal is
too small, less than 1\% even at large {\it IC} probability $w_{c}=$3.5\%.

In contrast, the $x$-spectra of $D{\bar D}$ and $D^-D^+$-mesons
calculated within the SM at large $x$ are almost insensitive to the IC
signal, because it is mostly determined by the diquark $x$-distributions
at large $x$, which do not take into account the {\it IC} effect.  The {\it extrinsic}
and {\it intrinsic}
sea quark distributions in the proton are also
suppressed compared to the diquark distribution at large $x$.
However, the {\it IC} signal can be visible at the asymmetry of
$D^-D^+$-mesons.
As mentioned in Sect. 4, the large diquark contributions at large $x$
are almost canceled in the difference $d\sigma_{D^-}/dx-d\sigma_{D^+}/dx$,
see Eq.(\ref{def:asymDplDmin}) for the $D^-D^+$ asymmetry. The {\it IC}
contributions are not canceled due to their asymmetrical $x$-dependence. 
This leads to an enhancement of the $D^-D^+$ asymmetry at large $x$.
  It is shown that the $D^+-D^-$ asymmetry increases if the intrinsic quark mass
  $M_Q$ decreases, as it can be seen in the left bottom of  Fig.~\ref{fig4}.
  Therefore, the contribution of the strange-antistrange intrinsic quarks ({\it IS})
  to the asymmetry of strange-antistrange hadrons  could be larger than the
  {\it IC} one
  to the $D^+D^-$ asymmetry. More detailed theoretical analysis of this effect can
  be done as
  the next step calculating {\it IS} distributions
  \cite{Peng:2012,BBLMST:2019} and
  taking into account different fragmentation functions of quarks to strange hadrons.

  There is another interesting and very important result. The asymmetry of $D^+D^-$
  and $D^0{\bar D}^0$ mesons changes the sign by the inclusion of the asymmetric  
  {\it IC} distribution instead of the symmetric one.


This paper thus confirms the important role of the $c{\bar c}$ asymmetry
for the {\it IC} content in the proton as 
obtained from lattice gauge theory in ref.~\cite{Sufian:2020coz} and
observations~\cite{NNPDF_Nov.2023} of the $c{\bar c}$ asymmetry from
$Z+c$ production in $pp$ collisions at the LHC ~\cite{LHCb:2022}.

\section{Acknowledgments}
We are very grateful to R.S.~Sufian for the data file of ${\Delta
  c(x)}=[c(x)-{\bar c}(x)]$ obtained within the lattice QCD which was
essential for
our calculations. We also thank V.A.~Bednyakov for useful discussions.  
This work was supported in part by 
the Dept.\ of Energy
Contract No.~DE-AC02-76SF00515. SLAC-PUB-17763.


\section{Appendix}
\subsection{Interference effects}

In the standard QGSM, the interference between different
topological graphs are not included. Let us try to take into account the
interference terms between different parton contributions,  including the 
contributions of extrinsic and intrinsic $c{\bar c}$ pairs to the inclusive
spectrum of $D$-mesons $\rho_h(x)$ presented by Eq.~\ref{def:rho_h}.
The relative phase between these contributions is unknown, therefore we can
estimate the maximum of this interference assuming the constructive sum
squared of the all amplitudes in the inclusive $x$-spectrum.
Thus, the inclusive $x$-spectrum of $D$-mesons:
\be
\label{equation:inclusiveCS}
\rho_D(x)=\int E\frac{d^3\sigma}{d^3p}d^2p_{\perp}=\sum\limits_{n=0}^{\infty}
\sigma_{n}(s)
\varphi_{n}^{D}(s,x),
\label{def:rho_D}
\ee
can be also presented in the form:
\begin{equation}
  \rho_D{tot}=\sum_{n=1}^{\infty}\sigma_{n}(s)\phi_{n}^{D}(s,x)~=~(\sqrt{A_1}~+~
  \sqrt{A_2})^2,
\label{def:rhotot}
\end{equation} 
where
\begin{equation}
A_1~=~a^D\sum_{n=1}^{\infty}\sigma_n \left\{F_{q_V}^{D(n)}(x_{+})F_{qq}^{D(n)}(x_{-}) +
F_{qq}^{D(n)}(x_{+})F_{q_V}^{D(n)}(x_{-})\right\}
\label{def:A1}
\end{equation}
and
\begin{equation}
  A_2~=~a^D\sum_{n=1}^{\infty}\sigma_n 2(n-1)F_{q_{\rm sea}}^{D(n)}(x_{+})
  F_{{\bar q}_{\rm sea}}^{D(n)}(x_{-}) 
\label{def:A2}
\end{equation}
were $A_2$ has both extrinsic and intrinsic $c{\bar c}$ contributions. 
We have calculated $A_2$ at 
$\omega_c=$ 0.0, 01, 0.035, and at different $c{\bar c}$
weights in the proton $\delta_c=$ 0.01 and $\delta_c=$ 0.05. The term $A_1$ does
not have both extrinsic and intrinsic $c{\bar c}$ contributions.

  
\subsection{Fragmentation functions  for $D^{+}$ meson}   

Let us present the fragmentation functions of quarks, diquarks into $D$-mesons 
(FF) used within the QGSM \cite{Lykasov:1999}.

Valence quark, diquark FF into $D^{+}$ meson:
\begin{equation}
	G_{d}^{D^{+}}(x) = G_{u}^{D^{+}}(x) = 
	(1-x)^{\lambda-\alpha_{\psi}(0)+2(1-\alpha_{R})
}
\end{equation}
\begin{equation}
	G_{uu}^{D^{+}}(x) = G_{ud}^{D^{+}}(x) = 
	(1-x)^{\lambda-\alpha_{\psi}(0)+2(\alpha_{R}-\alpha_{N})+1}
\end{equation}
Sea quark FF into $D^{+}$ meson:
\begin{equation}
	G_{\bar{u}}^{D^{+}}(x) = G_{u}^{D^{+}}(x) = G_{d}^{D^{+}}(x) =
	(1-x)^{\lambda-\alpha_{\psi}(0)+2(1-\alpha_{R})}
\end{equation}
\begin{equation}
	G_{\bar{d}}^{D^{+}}(x) = 
	(1-x)^{\lambda-\alpha_{\psi}(0)}(1+a_1{x}^2)
\end{equation}
\begin{equation}
	G_{{c}}^{D^{+}}(x) = G_{\bar{c}}^{D^{+}}(x) = 
	x^{1-\alpha_{\psi}(0)}(1-x)^{\lambda-\alpha_{R}}
\end{equation}
\subsection{Fragmentation functions  for $D^{-}$ meson}
Valence quark, diquark FF into $D^{-}$ meson:

\begin{equation}
	G_{u}^{D^{-}}(x) = 
	(1-x)^{\lambda-\alpha_{\psi}(0)+2(1-\alpha_{R})}
\end{equation}
\begin{equation}
	G_{d}^{D^{-}}(x) = 
	(1-x)^{\lambda-\alpha_{\psi}(0)}(1+a_1{x}^2)
\end{equation}
\begin{equation}
	G_{ud}^{D^{-}}(x) = 
	(1-x)^{\alpha_{R}-2\alpha_{N}+\lambda+\Delta\alpha+1}(1+(a_1{x}^2)/2)
\end{equation}
\begin{equation}
	G_{uu}^{D^{-}}(x) = 
	(1-x)^{\alpha_{R}-2\alpha_{N}+\lambda+\Delta\alpha+1}
\end{equation}
Sea quark FF into $D^{-}$ meson:
\begin{equation}
	G_{\bar{u}}^{D^{-}}(x) = G_{\bar{d}}^{D^{-}}(x) = G_{u}^{D^{-}}(x) =
	(1-x)^{\lambda-\alpha_{\psi}(0)+2(1-\alpha_{R})}
\end{equation}
\begin{equation}
	G_{d}^{D^{-}}(x) = 
	(1-x)^{\lambda-\alpha_{\psi}(0)}(1+a_1{x}^2)
\end{equation}
\begin{equation}
	G_{{c}}^{D^{-}}(x) = G_{\bar{c}}^{D^{-}}(x) = 
	x^{1-\alpha_{\psi}(0)}(1-x)^{\lambda-\alpha_{R}}
\end{equation}


\subsection{Fragmentation functions  for the $D^{0}$ meson}

Valence quark, diquark FF into $D^{0}$ meson:
\begin{equation}
	G_{d}^{D^{0}}(x) = G_{u}^{D^{0}}(x) = 
	(1-x)^{\lambda-\alpha_{\psi}(0)+2(1-\alpha_{R})}
\end{equation}
\begin{equation}
	G_{uu}^{D^{0}}(x) = G_{ud}^{D^{0}}(x) = 
	(1-x)^{\lambda-\alpha_{\psi}(0)+2(\alpha_{R}-\alpha_{N})+1}(1+a_1{x}^2)
\end{equation}
Sea quark FF into $D^{0}$ meson:
\begin{equation}
	G_{\bar{d}}^{D^{0}}(x) = G_{d}^{D^{0}}(x) = G_{{u}}^{D^{0}}(x) =
	(1-x)^{\lambda-\alpha_{\psi}(0)+2(1-\alpha_{R})}
\end{equation}
\begin{equation}
	G_{\bar{u}}^{D^{0}}(x) = 
	(1-x)^{\lambda-\alpha_{\psi}(0)}(1+a_1{x}^2)
\end{equation}
\begin{equation}
	G_{{c}}^{D^{0}}(x) = G_{\bar{c}}^{D^{0}}(x) = 
	x^{1-\alpha_{\psi}(0)}(1-x)^{\lambda-\alpha_{R}}
\end{equation}



\subsection{Fragmentation functions  for the ${\bar{D}^0}$ meson}

Valence quark, diquark FF into ${\bar{D}^0}$ meson:
\begin{equation}
	G_{d}^{\bar{D}^0}(x) = 
	(1-x)^{\lambda-\alpha_{\psi}(0)+2(1-\alpha_{R})}
\end{equation}
\begin{equation}
	G_{u}^{\bar{D}^0}(x) = 
	(1-x)^{\lambda-\alpha_{\psi}(0)}(1+a_1{x}^2)
\end{equation}
\begin{equation}
	G_{ud}^{\bar{D}^0}(x) = 
	(1-x)^{\alpha_{R}-2\alpha_{N}+\lambda+\Delta\alpha}
\end{equation}
\begin{equation}
	G_{uu}^{\bar{D}^0}(x) = 
	(1-x)^{\alpha_{R}-2\alpha_{N}+\lambda+\Delta\alpha}
\end{equation}
Sea quark FF into ${\bar{D}^0}$ meson:
\begin{equation}
	G_{\bar{d}}^{\bar{D}^0}(x) = G_{\bar{u}}^{\bar{D}^0}(x) = G_{d}^{\bar{D}^0}(x) =
	(1-x)^{\lambda-\alpha_{\psi}(0)+2(1-\alpha_{R})}
\end{equation}
\begin{equation}
	G_{u}^{\bar{D}^0}(x) = 
	(1-x)^{\lambda-\alpha_{\psi}(0)}
\end{equation}
\begin{equation}
	G_{{c}}^{\bar{D}^0}(x) = G_{\bar{c}}^{\bar{D}^0}(x) = 
	x^{1-\alpha_{\psi}(0)}(1-x)^{\lambda-\alpha_{R}}
\end{equation}

$\lambda=\alpha_{R}=0.5$; $\alpha_{N}=-0.5$; $\alpha_{\psi}(0)=-2.2$ or
$\alpha_{\psi}(0)=0$; $a_{1}=2$ or $a_{1}=30$; $\Delta\alpha=
\alpha_{R}-\alpha_{\psi}(0)$

\end{document}